\documentclass[12pt,english]{article}
\usepackage[a4paper, portrait,
	left=0.75in,
	right=0.75in,
	top=1in,
	bottom=0.5in,
	footskip=.25in]{geometry}
\usepackage[english]{babel}
\usepackage[utf8]{inputenc}
\usepackage[T1]{fontenc}
\usepackage{helvet,etoolbox,graphicx,titlesec}
\usepackage{caption,subcaption}
\usepackage[affil-it]{authblk} 
\usepackage{etoolbox}
\usepackage{lmodern}

\usepackage{mathrsfs}
\usepackage{amsfonts}
\usepackage{amsmath} 
\usepackage{amssymb} 

\usepackage{titlesec}

\usepackage{mathrsfs}

\usepackage{graphicx}
\usepackage{dcolumn}
\usepackage{bm}

\usepackage{amsthm}
\usepackage{xspace}
\usepackage{xfrac}
\usepackage[mathcal]{euscript}
\usepackage{tikz}
\usepackage{ifthen}

\usepackage{pgfplots}
\usepgfplotslibrary{patchplots,colormaps}
\pgfplotsset{width=7cm,compat=1.18,
colormap={mycolormap}{color=(black) color=(black!20!white)}}
\usepgfplotslibrary{fillbetween} 
\usepgflibrary{shadings}
\usetikzlibrary{backgrounds}
\usetikzlibrary{patterns}

\usepackage{epsdice}
\usepackage{twemojis}

\usepackage{pifont}

\usepackage{tcolorbox}
\tcbuselibrary{theorems,skins,breakable}
\usepackage{float}
\usepackage{colortbl}

\usepackage{natbib}

\usepackage{orcidlink}

\usepackage{url}
\usepackage{hyperref}
\usepackage{color}
\definecolor{refcolor}{RGB}{160,35,0}
\definecolor{hrefcolor}{RGB}{0,35,190}
\hypersetup{
    colorlinks,
    citecolor=refcolor,
    filecolor=refcolor,
    linkcolor=hrefcolor,
    urlcolor=hrefcolor
}

\usepackage{booktabs}
\usepackage{bookmark}
\bookmarksetup{
  numbered, 
  open,
}

\definecolor{greenPsi}{rgb}{0.0, 0.375, 0.0}
\definecolor{blueStruct}{rgb}{0.0, 0.0, 1.0}
\definecolor{redStruct}{rgb}{1.0, 0.0, 0.0}

\newcommand{\boxRule}{0.15mm}

\newcommand{\boxIndent}{15pt}

\newcommand{\remColor}{green}
\newcommand{\quoteColor}{green!50!yellow}
\newcommand{\questColor}{red}
\newcommand{\figColor}{orange}
\newcommand{\tabColor}{green}
\newcommand{\controlColor}{yellow}
\newcommand{\abstractColor}{blue!80!cyan}

\newenvironment{frameEnv}[1]
	{\begin{tcolorbox}[breakable,enhanced,toprule at break=0pt,bottomrule at break=0pt,before skip balanced=0.3cm,boxrule=\boxRule,left=0.75mm,right=0.75mm,frame hidden,borderline north = {\boxRule}{0pt}{#1!50!black}, borderline south = {\boxRule}{0pt}{#1!50!black},arc=0mm,colframe=#1!50!black,colback=#1!10,before upper={\parindent\boxIndent}]}
	{\end{tcolorbox}}

\newenvironment{frem}
	{\begin{frameEnv}{\remColor}}
	{\end{frameEnv}}
\newenvironment{fquest}
	{\begin{frameEnv}{\questColor}}
	{\end{frameEnv}}
\newenvironment{fquote}
	{\begin{frameEnv}{\quoteColor}}
	{\end{frameEnv}}
\newenvironment{ffig}
	{\begin{frameEnv}{\figColor}}
	{\end{frameEnv}}

	\newenvironment{frameEnvMargin}[1]
	{\begin{tcolorbox}[breakable,enhanced,toprule at break=0pt,bottomrule at break=0pt,before skip balanced=0.3cm,boxrule=\boxRule,left=0.75mm,right=0.75mm,top=5mm,bottom=5mm,frame hidden, borderline north = {\boxRule}{0pt}{#1!50!black}, borderline south = {\boxRule}{0pt}{#1!50!black},arc=0mm,colframe=#1!50!black,colback=#1!10,before upper={\parindent\boxIndent}]}
	{\end{tcolorbox}}

\newenvironment{fabstract}
	{\begin{frameEnvMargin}{\abstractColor}\begin{abstract}}
	{\end{abstract}\end{frameEnvMargin}}

\newtheorem{question}{Question}
\newtheorem{postulate}{Postulate}

\theoremstyle{remark}

\numberwithin{proofStep}{theorem} 

\theoremstyle{definition}
\newtheorem{defCustom}{Definition of}
\renewcommand{\thedefCustom}{\arabic{definition}}
\makeatletter
\newcommand{\setdefCustomtag}[1]{
  \let\oldthedefCustom\thedefCustom
  \renewcommand{\thedefCustom}{#1}
  \g@addto@macro\enddefCustom{
    \global\let\thedefCustom\oldthedefCustom}
  }
\makeatother

\theoremstyle{definition}
\newtheorem{observation}{Observation}

\newtheorem{condition}{Condition}
\renewcommand{\thecondition}{\arabic{condition}}
\makeatletter
\newcommand{\setconditiontag}[1]{
  \let\oldthecondition\thecondition
  \renewcommand{\thecondition}{#1}
  \g@addto@macro\endcondition{
    \global\let\thecondition\oldthecondition}
  }
\makeatother

\newtheorem{definition}{Definition}
\newtheorem{hypothesis}{Hypothesis}

\theoremstyle{remark}

\colorlet{disabled}{gray!50!black}

\newcommand{\orcid}[1]{\href{https://orcid.org/#1}{\textcolor[HTML]{A6CE39}{\aiOrcid}}}

\def\({\left(}
\def\){\right)}
\def\[{\left[}
\def\]{\right]}

\newcommand{\tn}{\textnormal}

\newcommand{\hilbert}{\mathcal{H}}

\newcommand{\mc}[1]{\mathcal{#1}}
\newcommand{\ms}[1]{\mathscr{#1}}

\newcommand{\wt}[1]{\widetilde{#1}}

\newcommand{\opern}[1]{\mathbf{#1}}
\newcommand{\oper}[1]{\hat{\opern{#1}}}

\newcommand{\R}{\mathbb{R}}
\newcommand{\C}{\mathbb{C}}

\newcommand{\abs}[1]{\lvert#1\rvert}

\newcommand{\ii}{\operatorname{i}}
\newcommand{\ee}{\operatorname{e}}

\newcommand{\ie}{\textit{i.e.}\xspace}

\newcommand{\etc}{\textit{etc}\xspace}

\newcommand{\schrod}{Schr\"odinger}

\newcommand{\bra}[1]{\langle#1|}
\newcommand{\ket}[1]{|#1\rangle}
\newcommand{\braket}[2]{\langle#1|#2\rangle}

\newcommand{\x}{\mathbf{x}}
\newcommand{\xThree}{\boldsymbol{x}}
\newcommand{\pThree}{\boldsymbol{p}}

\newcommand{\n}{\mathbf{n}}

\def\sref #1{\S\ref{#1}}

\usepackage{authblk}

\newsavebox\affbox
\author{Cristi Stoica\ \orcidlink{0000-0002-2765-1562}}
\affil{Dept. of Theoretical Physics, NIPNE---HH, Bucharest, Romania.\\
Email: \textit{\color{cyan}\href{mailto:cristi.stoica@theory.nipne.ro}{cristi.stoica@theory.nipne.ro},  \href{mailto:holotronix@gmail.com}{holotronix@gmail.com}}\vspace{-0.3in}}

\makeatletter
\newcommand*\@secondofsix[6]{#2}
\newcommand{\addtotitleformat}{%
  \@ifstar{\addtotitleformat@star}{\addtotitleformat@nostar}}
\newcommand\addtotitleformat@nostar[2]{%
  \PackageError{titlesec}{non starred form of \string\addtotitleformat\space not supported}{}}
\newcommand\addtotitleformat@star[2]{%
  \expandafter\expandafter\expandafter\expandafter
  \expandafter\expandafter\expandafter\def
  \expandafter\expandafter\expandafter\expandafter
  \expandafter\expandafter\expandafter\@currentsection@font
  \expandafter\expandafter\expandafter\expandafter
  \expandafter\expandafter\expandafter{%
    \expandafter\expandafter\expandafter\@secondofsix
       \csname ttlf@\expandafter\@gobble\string#1\endcsname}%
  \titleformat*{#1}{\@currentsection@font#2}%
}
\makeatother

\titlespacing\section{0pt}{20pt plus 6pt minus 4pt}{16pt plus 4pt minus 4pt}
\titlespacing\subsection{12pt}{12pt plus 6pt minus 4pt}{10pt plus 4pt minus 4pt}
\titlespacing\subsubsection{12pt}{12pt plus 6pt minus 4pt}{10pt plus 4pt minus 4pt}

\titleformat{\section}{\normalfont\fontsize{16}{24}\bfseries}{\thesection.}{1em}{}
\titleformat{\subsection}{\normalfont\fontsize{14}{20}\bfseries}{\thesubsection.}{1em}{}
\titleformat{\subsubsection}{\normalfont\fontsize{13}{18}\bfseries}{\thesubsubsection.}{1em}{}

\titleformat{\author}{\normalfont\fontsize{14}{20}\bfseries}{\thesection}{1em}{}

\makeatletter
\renewcommand{\thesubsection}{\arabic{section}.\arabic{subsection}}
\makeatother



\definecolor{titcolor}{RGB}{0,90,255}
\addtotitleformat*{\section}{\Large\sffamily\color{titcolor}}
\addtotitleformat*{\subsection}{\large\sffamily\color{titcolor}}
\addtotitleformat*{\subsubsection}{\large\sffamily\color{titcolor}}

\title{\color{titcolor}\textbf{{\fontsize{28.5}{40}{\bfseries Observation as Physication}}\\A single-world unitary no-conspiracy interpretation of quantum mechanics}}

\date{\small\today} 

\begin{document}

\pagestyle{headings}	
\newpage
\setcounter{page}{1}
\renewcommand{\thepage}{\arabic{page}}

\maketitle

\begin{fabstract}
The physical meaning of the operators is not reducible to the intrinsic relations of the quantum system, since unitary transformations can find other operators satisfying the exact same relations. The physical meaning is determined empirically. I propose that the assignment of physical meaning to operators spreads through observation, along with the values of the observables, from the already observed degrees of freedom to the newly observed ones. I call this process ``physication''. I propose that quantum observations are nothing more than this assignment, which can be done unitarily. This approach doesn't require collapse, many-worlds, or a conspiratorial fine tuning of the initial conditions.
\end{fabstract}



\begin{fquote}
[In] certain Indian theories of perception [...] the inner organ (\emph{anta\d{h}kara\d{n}a}) proceeds outward and illuminates the dark insentient world (like beams of light emitted from the eyes).
\begin{flushright}
Richard King, \emph{Early {A}dvaita {V}ed{\=a}nta and {B}uddhism} \citep{King1995EarlyAdvaitaVedantaAndBuddhism}.
\end{flushright}

So whenever the ray that flows through the eyes issues forth into surrounding daylight, like meets with like and coalesces with it, until a single, undifferentiated stuff is formed, in alignment with the direction of the eyes,
wherever the fire from inside strikes and pushes up against an external object. 
The similarity between the fire from within and the fire outside means that the stuff is completely homogeneous, and whenever it touches or is touched by anything else, it transmits the object’s impulses right through itself and all the way up to the soul,
and the result is the perception we call ‘seeing’.
\begin{flushright}
Plato, \emph{Timaeus}, sections 45c--45d \citep{PlatoWaterfield2008TimaeusAndCritias}.
\end{flushright}

There exist simultaneously two states, one being a predictive state $\Psi_p(t)$ which complies with the initial condition at $t=0$, and the other a retrodictive state $\Psi_r(t)$ which complies with the final condition at $t=\tau$. Both $\Psi_p(t)$ and $\Psi_r(t)$ obey the same Schr\"odinger equation\footnote{Here I propose that the correspondence between operators and physical properties spreads with each observation from the ``known'' to the ``unknown'', achieving the identity between $\Psi_p(t)$ and $\Psi_r(t)$.}. 
\begin{flushright}
\citep{Watanabe1955SymmetryOfPhysicalLawsPartIIIPredictionAndRetrodiction}. Also see \citep{AharonovBergmannLebowitz1964TimeSymmetryInTheQuantumProcessOfMeasurement}.
\end{flushright}
\end{fquote}

\section{Introduction}
\label{s:intro}

This article discusses two problems, and suggests that they have a common explanation. One is the well-known measurement problem, which is due to an apparent contradiction between the evolution law of a quantum system, and the fact that quantum measurements have definite results \citep{vonNeumann1955MathFoundationsQM}.
The second problem, much more recent \citep{Stoica2022SpaceThePreferredBasisCannotUniquelyEmergeFromTheQuantumStructure,Stoica2023AreObserversReducibleToStructures}, is that the relational, structural properties of the stuff making the world are unable to provide a unique physical meaning.
The first problem shows an overdetermination, and the second problem exposes an underdetermination. This suggests that these problems are complementary and solving the former also solves the latter.

Previously, it was shown that there is no way to determine what physical property an operator represents, only from relational or structural considerations \citep{Stoica2022SpaceThePreferredBasisCannotUniquelyEmergeFromTheQuantumStructure,Stoica2023AreObserversReducibleToStructures}. This huge underdetermination of the physical meaning by the structure can be resolved only by observation.
I propose that quantum observation does not ``measure'' or ``collapse'' the state of the observed system, but rather,  without disturbing the state by changing the dynamics or by inserting \emph{ad hoc} projections or by ignoring branches of the wavefunction, it simply assigns physical properties to the operators. This transfer of physical meaning from the observer and the measuring device to the observed system is called here ``physication''.

This leads directly to an interpretation of quantum mechanics with the following features:

\begin{enumerate}
	\item 
no modification of the {\schrod} dynamics, no additional entities,
	\item 
evolution is always unitary, without collapse or branching,
	\item 
measurements have definite outcomes, so there is always a single world,
	\item 
physication is nondeterministic,
	\item 
the wavefunction is ontic, space emerges through physication,
	\item 
with respect to the resulting space, all interactions are local.
\end{enumerate}

The idea is very simple. Consider for example the case of non-relativistic quantum mechanics, where the Hamiltonian operator is a function of the position and the momentum operators, as well as spin and internal degrees of freedom. But from a relational or structural point of view, these operators can be chosen in infinitely many ways and still satisfy the exact same relations, and the Hamiltonian operator has the same functional dependency in terms of them (\citealp{Stoica2022SpaceThePreferredBasisCannotUniquelyEmergeFromTheQuantumStructure}, \S 2.2).
Then, if we are interested in the position of the observed system, for a given abstract state vector and a Hamiltonian, the symmetry reveals that there is no preferred choice of what basis should correspond to positions.
No matter what invariant conditions we impose, for example that the Hamiltonian is local in that basis, there are infinitely many bases satisfying the same conditions.
 I will say more about this in Section \sref{s:underdetermination}.

Then, we can interpret quantum observations as determining not only the particle's position, but also the position basis of the observed system's state space. Thus, while the state vector evolves unitarily, the position basis of the universe emerges gradually, being extended to other systems through quantum observations, a process that we call \emph{physication}. The freedom given by unitary symmetry makes this process to be random, even if the evolution is deterministic (for being unitary). Since the physication of macroscopic systems is never complete, there is a reservoir of unphysicated degrees of freedom in the preparation and the measuring apparatus. This allows the process to be applied for successive noncommutting measurements, without invoking the wavefunction collapse or branching. If we insist to assume that the position basis is predetermined, because the measurements have definite outcomes without collapse or branching, successive measurements appear as a conspiracy between the initial conditions of the observed system and the experimental setup, but everything is local. If we assume that the position basis is not predetermined, then physication is understood as localization, so that it makes no sense to talk about a position prior to localization.

The problem of underdetermination of the operators representing each physical property is demonstrated in Section \sref{s:underdetermination}.
The complementary problem, that quantum measurement is a problem of overdetermination, is discussed in Section \sref{s:overdetermination}.
The proposed interpretation of quantum mechanics, based on physication, and the way it avoids the contradiction implied by the measurement problem, are presented in Section \sref{s:physication}.
A formulation in terms of Postulates is given in Section \sref{s:physication-postulates}.
The article ends with a discussion of the possibility of empirical falsification, in Section \sref{s:experiment}, and with a comparison with other proposals, in Section \sref{s:comparison}.

\section{The underdetermination of the physical properties}
\label{s:underdetermination}

Can everything that can be said about the world be reduced to structures and their purely relational properties? The successes of modern physics, especially its three pillars, relativity, quantum theory, and gauge theory, taught us that this should indeed be the case. They taught us about the power of symmetries, invariance, representations, relations, and how to understand them.
Special relativity taught us to give up the idea that there is a kind of material through which the electromagnetic waves propagate. General relativity taught us that spacetime itself behaves like a field that we call gravity. Gauge theory pushed the geometrization of matter even more, so that similar curvature effects as in General Relativity take place in some internal dimensions, observable only indirectly, due to their effects, which are the gauge forces. Quantum mechanics taught us that whatever propagates, it appears to propagate in the abstract, high-dimensional configuration space, rather than in three dimensions. Quantum field theory introduced even more abstractions and alternative ways to describe reality.
Since the same phenomena can be described in different ways, it would be all the same, we think, if we use one way or another to describe the world, as long as these descriptions translate into one another biunivocally. We can use the Heisenberg picture, the Schr\"odinger picture, or the interaction picture. We can use the phase space or the configuration space. And if the AdS/CFT correspondence \citep{Maldacena1999AdSCFTInitial} were correct, it would be all the same whether we prefer the gauge aspect or the gravity aspect of this correspondence, even if they live in spaces with different numbers of dimensions \citep{Wallace2022StatingStructuralRealismMathematicsFirstApproachesToPhysicsAndMetaphysics}. All these theories seem to teach us the same lesson, that all that matters is the relations between the entities. These relations provide the appropriate description of the physical reality, and moreover, as long as the constituting elements of different descriptions can be put in a one-to-one correspondence, they are equivalent.

So our theoretical models are based on the idea that all that matters is the structures formed by these relations.
Our theories admit mathematical descriptions precisely because they are structural and relational \citep{ChangAndKeisler1990ModelTheory,Hodges1997ShorterModelTheory,Gratzer2008UniversalAlgebra}. But this is equally true of experiments, since whenever we can measure something, we do it by comparing with some standard unit, which is another object, so again, all we can access are relations, out of which we build structures.
It is tempting to ``visualize'' the world as made of more familiar materials, but with modern physics, we gave up this circular ontology, especially when we realized that our previous intuitions were misleading, so now all that matters to a physicist is the structure alone, the relations, but not the relata, the stuff having that structure. The following hypothesis became an inherent, although usually unacknowledged explicitly, axiom of our modern theories:
\begin{fquest}
\begin{hypothesis}[Structural realism]
\label{hypothesis:SR}
Everything that can be known of the physical world, and our descriptions of the world, are nothing but structures.
\end{hypothesis}
\end{fquest}

This understanding strongly suggested to us that, in quantum mechanics, the operators can tell us everything there is to know about the physical properties they represent.

Consider for example the following Hamiltonian operator for $\n$ particles in non-relativistic quantum mechanics, expressed in function of the position and the momentum operators:
\begin{equation}
\label{eq:schrod_hamiltonian_NRQM}
\oper{H} = \sum_{j=1}^\n\frac{1}{2m_j}\hat{\pThree}_{j}^2
+ \mathop{\sum_{j=1}^\n\sum_{k=1}^\n}_{j\neq k}\hat{V}_{j,k}\(\abs{\hat{\xThree}_j-\hat{\xThree}_k}\),
\end{equation}
where 
$m_j$ is the mass of the particle $j$,
\begin{equation}
\label{eq:position-momentum}
\begin{aligned}
\hat{\xThree}_{j}&=\(\hat{x}_{j1},\hat{x}_{j2},\hat{x}_{j3}\)\\
\hat{\pThree}_{j}&=\(\hat{p}_{j1},\hat{p}_{j2},\hat{p}_{j3}\)\\
\end{aligned}
\end{equation}
are the position, respectively the momentum operators for the particle $j$, $\hat{\pThree}_{j}^2=\hat{p}_{j1}^2+\hat{p}_{j2}^2+\hat{p}_{j3}^2$, and the potential $\hat{V}_{j,k}$ depends only on the absolute value of the differences of the positions, $\abs{\hat{\xThree}_j-\hat{\xThree}_k}=\sqrt{\(\hat{x}_{j1}-\hat{x}_{k1}\)^2+\(\hat{x}_{j2}-\hat{x}_{k2}\)^2+\(\hat{x}_{j3}-\hat{x}_{k3}\)^2}$.

A unitary transformation with the operator $\oper{S}$ results in new operators,
\begin{equation}
\label{eq:position-momentum-transformed}
\begin{array}{rll}
\hat{\xThree}'_{j}&=\(\hat{x}'_{j1},\hat{x}'_{j2},\hat{x}'_{j3}\)&:=\(\oper{S}\hat{x}_{j1}\oper{S}^\dagger,\oper{S}\hat{x}_{j2}\oper{S}^\dagger,\oper{S}\hat{x}_{j3}\oper{S}^\dagger\)\\
\hat{\pThree}'_{j}&=\(\hat{p}'_{j1},\hat{p}'_{j2},\hat{p}'_{j3}\)&:=\(\oper{S}\hat{p}_{j1}\oper{S}^\dagger,\oper{S}\hat{p}_{j2}\oper{S}^\dagger,\oper{S}\hat{p}_{j3}\oper{S}^\dagger\).\\
\end{array}
\end{equation}

These operators have the exact same spectra and stand with respect to each other in the exact same relation as their original counterparts, that is, they, like the original position and momentum operators, satisfy the \emph{canonical commutation relations} (CCR):
\begin{equation}
\label{eq:CCR-prime}
\[\hat{x}'_{j\alpha},\hat{p}'_{k\beta}\]=\ii\hbar\delta_{j k}\delta_{\alpha\beta}\oper{I}.
\end{equation}

Therefore, we will call the operators $(\hat{\xThree}'_{j})_{j\in\{1,\ldots,\n\}}$ \emph{position-like operators}. We will call their common eigenbasis \emph{position-like basis} and label its vectors based on the eigenvalues, $\ket{\xThree'_1,\xThree'_2,\ldots,\xThree'_\n}$.

Since there are infinitely many transformations $\oper{S}$, how do we know if the operators \eqref{eq:position-momentum}, or any of the pairs of operators \eqref{eq:position-momentum-transformed}, each of them satisfying the CCR \eqref{eq:CCR-prime}, are the ``true'' position and momentum operators?
Recall that the idea is to choose them based on relations, which should consist of invariant identities or inequalities, that is, these relations should not change their form when expressed in different bases, just like the CCR \eqref{eq:CCR-prime}.
For example, we can choose those pairs of operators that make the Hamiltonian be local when expressed in function of these operators, as it is in equation \eqref{eq:schrod_hamiltonian_NRQM}. To be sure, let's just impose the condition that the Hamiltonian has the exact same form as \eqref{eq:schrod_hamiltonian_NRQM},
\begin{equation}
\label{eq:schrod_hamiltonian_NRQM-prime}
\oper{H} = \sum_{j=1}^\n\frac{1}{2m_j}\hat{\pThree}'_{j}{}^2
+ \mathop{\sum_{j=1}^\n\sum_{k=1}^\n}_{j\neq k}\hat{V}_{j,k}\(\abs{\hat{\xThree}'_j-\hat{\xThree}'_k}\).
\end{equation}

This Hamiltonian is local, and moreover, it has the exact same dependence of the position and momentum operators as \eqref{eq:schrod_hamiltonian_NRQM}. Does this condition, combined with the CCR, ensure the uniqueness of the position and the momentum operators?
It certainly doesn't, as we can see if we choose the unitary transformation $\oper{S}$ from equation \eqref{eq:position-momentum-transformed} to commute with $\oper{H}$, that is,
\begin{equation}
\label{eq:preserve-Hamiltonian}
\oper{H}=\oper{S}\oper{H}\oper{S}^\dagger.
\end{equation}

Then, the functional dependency \eqref{eq:schrod_hamiltonian_NRQM} of $\oper{H}$ of the operators $(\hat{\xThree}_{j})_{j\in\{1,\ldots,\n\}}$ and $(\hat{\pThree}_{j})_{j\in\{1,\ldots,\n\}}$, and its functional dependency \eqref{eq:schrod_hamiltonian_NRQM-prime} 
of the operators $(\hat{\xThree}'_{j})_{j\in\{1,\ldots,\n\}}$ and $(\hat{\pThree}'_{j})_{j\in\{1,\ldots,\n\}}$ from equation \eqref{eq:position-momentum-transformed}, are the same, regardless of the choice of the unitary transformation $\oper{S}$ commuting with $\oper{H}$.

There are infinitely many ways to choose such a transformation.
An obvious example is to choose transformations of the form $\oper{S}:=\ee^{-\frac{\ii}{\hbar}\oper{H}s}$, for any $s\in\R$.
The Hamiltonian doesn't commute with the position operators (a wavefunction concentrated at a definite position spreads immediately), so in general these transformations result in $\hat{\xThree}'_{j}\neq\hat{\xThree}_{j}$.

But choosing as the generator of our transformations $\oper{S}$ leaving the Hamiltonian unchanged the operator $\oper{H}$ itself this is only a way, there are infinitely many others.
For example, even in those systems with a two-dimensional state space, for whatever Hamiltonian there are infinitely many ways to choose a transformation $\oper{S}$ so that equation \eqref{eq:preserve-Hamiltonian} holds. Let's check this in a basis that diagonalizes the Hamiltonian:
\begin{equation}
\label{eq:reserve-Hamiltonian-two}
\oper{H}=
\begin{pmatrix}
a & 0 \\
0 & b
\end{pmatrix}.
\end{equation}

If $a=b$, any unitary operator $\oper{S}$ commutes with $\oper{H}$.
If $a\neq b$, any operator of the form
\begin{equation}
\label{eq:reserve-Hamiltonian-two-commutes}
\oper{S}=\begin{pmatrix}
\ee^{-\ii c} & 0 \\
0 & \ee^{-\ii d}
\end{pmatrix},
\end{equation}
where $c,d\in\R$, is unitary and commutes with $\oper{H}$.
In general, if the state space has $d$ dimensions, the unitary operators that commute with the Hamiltonian form a manifold of real dimension between $d$ and $d^2$. For an infinite-dimensional state space, the manifold of unitary operators that commute with $\oper{H}$ has an infinite number of dimensions.

Therefore, this ambiguity is extremely large. One may hope that we can reduce it by imposing more conditions, but as long as these conditions are expressed by invariant or basis-independent relations, they will be preserved by such transformations, so the manifold of possible transformations remains the same.

Despite this simple observation, there are several research programs that aim to recover the physical meanings of the operators, the three-dimensional space, the decomposition into subsystems, the preferred pointer states, classicality \etc, only from the Hamiltonian $\oper{H}$ and sometimes from a unit vector $\ket{\psi}$ representing the state of the system \citep{Piazza2010GlimmersOfAPreGeometricPerspective,Tegmark2015ConsciousnessAsAStateOfMatter,CotlerEtAl2019LocalityFromSpectrum,ZanardiDallasLloyd2022OperationalQuantumMereologyAndMinimalScrambling,Deutsch1985QTasUniversalPhysicalTheory,CarrollSingh2019MadDogEverettianism,Carroll2021RealityAsAVectorInHilbertSpace,CarrollSingh2021QuantumMereology}.
But the only structures that can be obtained uniquely are those with respect to which the state exhibits no physical change in time, \ie operators or structures that can be built only out of the spectrum and the eigenspaces of the Hamiltonian, and none of these programs are about the emergence of such quite trivial structures.
In \citep{Stoica2022SpaceThePreferredBasisCannotUniquelyEmergeFromTheQuantumStructure} it was proved that these programs fail to achieve a unique result, whatever kinds of conditions they may want to impose to the operators or the structures they want to obtain. More specific proofs and examples were given in \citep{Stoica2023PrinceAndPauperQuantumParadoxHilbertSpaceFundamentalism,Stoica2024DoesTheHamiltonianDetermineTheTPSAndThe3dSpace}.

Let's summarize this:
\begin{fquest}
\begin{observation}[Incompleteness of structural realism]
\label{obs:SR-ambiguity}
Only from structures and relations, there is no way to know unambiguously what physical property an operator represents.
\end{observation}
\end{fquest}

One may think that this is a false problem, since physics is basis independent. Physics is indeed basis independent, but here I am not talking about expressing the same physical phenomena in different bases. Such a change of basis would change the state vector too, so the state vector and the operators will be in the same relations regardless of the basis, including the mean values, so the physics described will be the same. Here I am talking about something completely different: identifying what operator represents each physical property based solely on the relational properties. And we've seen that there is no way to do this.

But what can solve this ambiguity? How do we, the observers, know what physical properties are out there and correspond to the operators? A tempting answer would be ``we know it from experiments!''. Let's consider this possibility. For an experiment, we need observers doing the experiments. So we need to understand what an observer is. Are the observers themselves characterized only by their structures and the dynamics of these structures, how they change in time? The materialistic theories of mind would say that structure is all that needs to be taken into account, for example that consciousness itself is nothing but a computation, as in the \emph{computational theory of mind} \citep{sep-computational-mind}, or maybe it consists of more sophisticated functionalities of the physical structures, as in \emph{functionalism} \citep{sep-functionalism}. In no way would a materialistic theory of mind accept that something beyond the structure is needed for consciousness, since this would allow for the in principle possibility that two structurally identical systems can be so that one is conscious and the other is not, depending on the stuff they're made of. That is, it would allow for the possibility of \emph{philosophical zombies} -- structures that have no consciousness at all, despite having the same structure and behaving in the same way as a conscious being \citep{KirkSquires1974ZombiesVMaterialists,sep-zombies}.
More about this is explained in \citep{Stoica2023AreObserversReducibleToStructures,Stoica2023AskingPhysicsAboutPhysicalismZombiesAndConsciousness}.

Suppose then that we know exactly what kind of structure can be an observer. Let's call such a structure \emph{observer-like structure} \citep{Stoica2023AreObserversReducibleToStructures}. This structure should be encoded in the patterns of the wavefunction. But the wavefunction $\ket{\psi(\xThree_1,\xThree_2,\ldots,\xThree_\n,t)}$ consists of the components of the state vector in the position basis, $\ket{\psi(\xThree_1,\xThree_2,\ldots,\xThree_\n,t)}=\braket{\xThree_1,\xThree_2,\ldots,\xThree_\n}{\psi(t)}$ (ignoring the spin and internal degrees of freedom here, for simplicity). And since the position operators are not uniquely defined by structures, how would we know the position basis? Wouldn't the operators $\hat{\xThree}'_{j}$  as in equation \eqref{eq:position-momentum-transformed} give a different wavefunction for the same state vector, with different patterns, and hence different structures that may or may not be conscious observers?
It surely would, since the wavefunction $\ket{\psi(\xThree_1,\xThree_2,\ldots,\xThree_\n,t)}$ looks completely different from the wavefunction $\ket{\psi(\xThree'_1,\xThree'_2,\ldots,\xThree'_\n,t)}$, in fact any state vector $\ket{\psi(t)}$ can look like any wavefunction $\ket{\wt{\psi}(\xThree'_1,\xThree'_2,\ldots,\xThree'_\n,t)}$, if expressed in a conveniently chosen basis. This is obvious if we remember that the unitary transformations can take any unit vector into any other unit vector. Therefore, any wavefunction $\ket{\psi(\xThree_1,\xThree_2,\ldots,\xThree_\n,t)}$ can appear like a completely different wavefunction $\ket{\psi(\xThree'_1,\xThree'_2,\ldots,\xThree'_\n,t)}$ for some position-like operators $(\hat{\xThree}'_{j})_{j\in\{1,\ldots,\n\}}$ obtained as in equation \eqref{eq:position-momentum-transformed}.

And an observer-like structure with respect to the operators $(\hat{\xThree}'_{j})_{j\in\{1,\ldots,\n\}}$ would disagree with an observer-like structure with respect to the operators $(\hat{\xThree}_{j})_{j\in\{1,\ldots,\n\}}$ about what operators represent the positions.

Note that, based on the example of Hamiltonian from equation \eqref{eq:schrod_hamiltonian_NRQM}, so far I mentioned only position-like operators and bases. In reality, in addition to the position degrees of freedom, there are spin and internal degrees of freedom. Taking these into account doesn't change the conclusion, in fact it increases even more the range of possibilities to assign operators to the physical properties.

An observer-like structure will always determine as position operators the operators with respect to which it appears as observer-like structure.
\begin{fquest}
\begin{observation}[Relativity of structure]
\label{obs:relativity-of-structure}
The observer-like structures cannot determine an absolute correspondence between the operators and the physical properties these operators are supposed to represent, based on structure and relations alone.
\end{observation}
\end{fquest}

Structural realism implies that all assignments of operators to physical properties are equally valid, provided that the operators are unitarily equivalent as, for example, in equation \eqref{eq:position-momentum-transformed}. Then, if consciousness also reduces to structures, this implies that all observer-like structures existing simultaneously as patterns of the wavefunctions expressing the same state vector in different position-like bases should be equally conscious. This means that we could be any of these observer-like structures. But it turns out that this can't be the case.
If an observer knows the value of a property of an external object, there are infinitely many observer-like structures that are identical with the observer, that contain in their brain-like structures the same information about the external world, but whose external world contains in its structure a version of the external object having different values for its property, as shown elsewhere \citep{Stoica2023AreObserversReducibleToStructures}.
It was also shown that this implies that the correlation between the records in the observer-like structures and the properties of the external objects is zero \citep{Stoica2023AreObserversReducibleToStructures}. In other words, observer-like structures know nothing about the external world. And this is true even if we restrict the transformations $\oper{S}$ to those that commute with the Hamiltonian, so that the wavefunctions in which the observer-like structures are patterns propagate according to a Schr\"odinger equation of the same form.
\begin{fquest}
\begin{observation}[Absoluteness of meaning]
\label{obs:absoluteness-of-meaning}
The correspondence between operators and physical properties has to be unique, otherwise we would know nothing about the external world.
\end{observation}
\end{fquest}

But this of course invalidates structural realism. Moreover, it requires that there are some special kind of observer-like structures that are sentient, while all the others, even if they have the same structure and evolve in the same way, are philosophical zombies.

\begin{fquest}
\begin{observation}[Absoluteness of observers]
\label{obs:zombies}
Only the observer-like structures with a particular assignment of physical properties to the operators are conscious observers.
\end{observation}
\end{fquest}

Details can be found in \citep{Stoica2023AreObserversReducibleToStructures}.
To understand this, consider the hypothesis that we can simulate conscious observers by simulating the wavefunction containing them.
This is in line with structural realism, and it is consistent with the theories of mind based on reduction to structure, like \emph{computationalism} \citep{sep-computational-mind} and \emph{functionalism} \citep{sep-functionalism}.
The simulation of the wavefunction can be understood as simulating the position operators by other operators, carefully selected to have the same spectrum, and so that the simulated wavefunction evolves, under the Hamiltonian of the system simulating it, in the same way as the original wavefunction.
But this is exactly what I discussed above, that, based on structure alone, there is no unique choice of the position operators.
This means that such simulations already happen due to the inherent ambiguity of the relations, but we don't see them, because our wavefunction contains them in an ``encrypted'' form under a unitary transformation. This ``encryption'' is only relative, being just a reinterpretation of the assignment of operators to represent physical properties.
If a simulation of an observer is as conscious as the observer it simulates, then all these coexisting simulations should support conscious observers. But, as shown in \citep{Stoica2023AreObserversReducibleToStructures}, this contradicts our observations, by implying that there would be no correlation between the values of the properties of the world and our own thoughts about the values of these properties.

Now, one may try to reject this proof based on correlation by claiming that, just because the reader happens to have a good correlation between their representation of the world and the world itself, it doesn't mean that the other observer-like structures are not observers.
But for each observer who has such a good correlation, there are infinitely many observer-like structures, in the eigenbases of other position-like operators $(\hat{\xThree}'_{j})_{j\in\{1,\ldots,\n\}}$, and for these observer-like structures, their external world is nothing as they think it is. Such observer-like structures are akin to the Boltzmann brains \citep{Eddington1934NewPathwaysInScience_Boltzmann_brains}: even if in the present they are structurally identical to the original observer, very soon they would experience a very unpredictable world, they would even disintegrate.
Note that the observer-like structures encoding the wrong information about the properties of the external worlds are ``like'' the Boltzmann brains, but this is not the same situation. Standard Boltzmann brains don't need a reassignment of the operators representing the physical properties in order to exist, they are due to fluctuations.
In our case, infinitely many such observer-like structures exist whenever an observer exists, but they are ``hidden'' in other assignments of operators to the physical properties.
So, if the reader doesn't experience such unfortunate events, either they are extremely lucky, or simply the other observer-like structures can't support sentience, even if they have the right structure for this, and so the reader can't be such an observer-like structure.
The second option violates structural realism.
This is the meaning of the result from \citep{Stoica2023AreObserversReducibleToStructures}.

But now, we notice that it is not sufficient that the choice of the position operators is unique (in spite of the fact that the structure alone doesn't single out a unique choice).
The reason is that this choice should also prevent the real, conscious observers, from being observer-like structures that are akin to the Boltzmann brains, that is, having mental representations unrelated with the properties of the external world. In other words
\begin{fquest}
\begin{observation}[The past hypothesis]
\label{obs:past-hypothesis}
The choice of the position operators is such that the initial state of the wavefunction in the position configuration space is constrained in the exact way that ensures the existence of the arrow of time.
\end{observation}
\end{fquest}

Without this very strong constraint, the universe would not be predictable and the memories of the observers would not be correct.
It is presumed that the initial state had a very low entropy, so that the Second Law of Thermodynamics is ensured \citep{Boltzmann1910VorlesungenUberGastheorie,Feynman1965TheCharacterOfPhysicalLaw}.
In fact, this constraint is much stronger than we used to think based on the increase of entropy alone \citep{Stoica2024DoesQuantumMechanicsRequireConspiracy}.

But now, given all these, the following question arises:
\begin{fquest}
\begin{question}[Unfinished operators-physical properties correspondence]
\label{question:delayed-correspondence}
Is the choice of the position operators God-given at the Big Bang, or is it happening as we make new observations?
\end{question}
\end{fquest}

Given Observation \ref{obs:past-hypothesis}, if this correspondence was completely given since the beginning, it seems to be extremely fine-tuned \citep{Stoica2024DoesQuantumMechanicsRequireConspiracy}.
This justifies looking into the possibility that it is not finalized, but it's still happening. In this article, we will explore this possibility, and see that it can provide a new kind solution to the measurement problem, a solution that avoids breaking the Schr\"odinger equation by collapse, but also avoids the split into many worlds, and at the same time doesn't put the burden of this avoidance on a conspiratorial choice of the initial conditions of the universe.

\section{The overdetermination of the quantum states}
\label{s:overdetermination}

The \emph{measurement problem} is a problem of \emph{overdetermination}, because it seems to require the state vector to satisfy contradictory conditions.
On the one hand, the evolution law, the Schr\"odinger equation, predicts a certain future state from a present state. On the other hand, for quantum measurements, the evolution of the state seems to lead to a superposition of distinct outcomes. The projection postulate is invoked to leave us with a definite outcome, in conflict with the evolution law \citep{vonNeumann1955MathFoundationsQM,Dirac1958ThePrinciplesOfQuantumMechanics}. That is, to resolve the tension between the evolution law and the results of quantum measurements, the standard choice is to sacrifice the evolution law, so that the state jumps into the state revealed by the observation.

\begin{fquest}
\begin{observation}[Overdetermination]
\label{obs:overdetermination}
There seems to be a tension between the evolution law and the projection postulate, resulting in an overdetermination of the quantum states.
\end{observation}
\end{fquest}

Moreover, assuming that the property to be measured already has a definite value waiting to be measured, the collapse seems to happen non-locally \citep{EPR35,Bohm1951TheParadoxOfEinsteinRosenAndPodolsky,Bell64BellTheorem}. 
For someone who wants a compelling mechanism to explain what we see in a quantum world, this non-locality seems to be the way to go \citep{Bohm1952SuggestedInterpretationOfQuantumMechanicsInTermsOfHiddenVariables,GhirardiRiminiWeber1986GRWInterpretation,Bell2004SpeakableUnspeakable}.
It may be at odds with the relativity of simultaneity \citep{einstein1905elektrodynamik}, but it can't be used for faster than light signaling \citep{GhirardiRiminiWeber1980AGeneralArgumentAgainstSuperluminalTransmissionThroughQuantumMechanicalMeasurementProcess,PeresTerno2004QuantumInformationAndRelativityTheory}.
However, if we care to have a compelling description of the quantum phenomena, beyond the instrumentalist slogan ``shut up and calculate'', shouldn't we not be instrumentalists also about relativity?
A wavefunction collapse, but also a non-local mechanism, would violate the relativity of simultaneity, not at the empirical level, but at the fundamental level.

Collapse would also violate the conservation laws. And, while in the many-worlds interpretation (MWI) \citep{Everett1957RelativeStateFormulationOfQuantumMechanics} and in the pilot-wave theory (PWT) \citep{Bohm1952SuggestedInterpretationOfQuantumMechanicsInTermsOfHiddenVariables} there is no collapse to break the evolution law, per branch the conservation laws would still be violated \citep{Burgos1993ConservationLawsQM,Stoica2017TheUniverseRemembersNoWavefunctionCollapse,Stoica2021PostDeterminedBlockUniverse}.
For all these interpretations, this happens even if we take into account the possibility of conservation for the combined system, consisting of the measuring device and the observed system.
The only way out seems to be to have a single-world unitary theory, without collapse, \citep{Stoica2008SmoothQuantumMechanics}, but this would require conspiratorial initial conditions (\citealp{Stoica2012QMQuantumMeasurementAndInitialConditions}, also because of \citealp{Bell64BellTheorem,KochenSpecker1967HiddenVariables}).

At any rate, wouldn't it be great if we would have a single-world unitary evolution without collapse and without branching? This would avoid at once the violation of the evolution law, the violation of the conservation laws, and it would be consistent with relativity \citep{Stoica2008SmoothQuantumMechanics}.
Attempts to resolve the measurement problem by single-world unitary evolution, that is, without collapse and without branching, can be traced back to the works of Schulman \citep{Schulman1984DefiniteMeasurementsAndDeterministicQuantumEvolution,Schulman2017ProgramSpecialState}, 't Hooft \citep{tHooft2016CellularAutomatonInterpretationQM} and Stoica \citep{Stoica2008SmoothQuantumMechanics,Stoica2012QMGlobalAndLocalAspectsOfCausalityInQuantumMechanics,Stoica2016OnTheWavefunctionCollapse,Stoica2017TheUniverseRemembersNoWavefunctionCollapse,Stoica2021PostDeterminedBlockUniverse}.

Attempts to make sense of this consisted of using the block universe picture drawn to us by relativity, and distributing the initial conditions so that they are not given in their final form at the Big Bang, but, for some yet unobserved degrees of freedom, they are delayed in various places and times across the block universe \citep{Stoica2008SmoothQuantumMechanics,Stoica2012QMGlobalAndLocalAspectsOfCausalityInQuantumMechanics,Stoica2021PostDeterminedBlockUniverse}. Each observation is local, and the solutions compatible with it are determined locally. The global solution has to be obtained by extending local solutions, as in \emph{sheaf theory} \citep{MacLaneMoerdijk1992SheavesInGeometryAndLogic,Bredon1997SheafTheory}.
So each new observation reduces the set of global solutions compatible with all observations done up to that moment.
This should be possible, because most degrees of freedom are unavailable at the macroscopic level, and this may leave room for the state to be determined with a delay in a way that is consistent with all of the observations.
However, from the perspective of a description in which the initial conditions are fixed from the beginning, this would still appear as conspiratorial \citep{Stoica2012QMQuantumMeasurementAndInitialConditions}.
But now, in the light of Section \sref{s:underdetermination}, we have the opportunity to reconsider this apparent conspiracy.
\begin{fquest}
\begin{question}
\label{obs:reevaluation}
Is it possible that the underdetermination of the physical properties by the structure compensates for the apparent overdetermination of the quantum state, resulting in a single-world unitary interpretation that is not conspiratorial?
\end{question}
\end{fquest}

Let's explore this new avenue.

\section{Observation as physication}
\label{s:physication}

In this Section I apply the idea of physication to show how the measurement problem does not require violations of unitary evolution in a single world.
In the next Section I will propose the Postulates of this new interpretation of quantum mechanics.

The simplest case is the observation of a particle that was never observed so far. Suppose it is a spin-$1/2$ particle, and we measure its spin along the $z$-axis.
If the particle's spin is in a pure state, it is represented by a vector in a two-dimensional complex space. Let $\(\ket{\alpha},\ket{\beta}\)$ be an orthonormal basis of this space, and let the state vector be $\ket{v}=a\ket{\alpha}+b\ket{\beta}$, where $a^\ast a + b^\ast b=1$.

The operator representing the particle's spin along the $z$-axis, let us denote it by $\oper{S}_z$, is not yet assigned, so it can be any Hermitian operator with the spectrum $\{-\frac 1 2\hbar,+\frac 1 2\hbar\}$.
There are infinitely many such operators, and therefore there are infinitely many unitary transformations of the spin's space that can map the basis $\(\ket{\alpha},\ket{\beta}\)$ into the eigenbasis of such an operator.
Suppose we measured the spin along the $z$-axis and obtained the eigenvalue $+\frac 1 2\hbar$.
Then, we can choose 
\begin{equation}
\label{eq:sigma-z-choice-up}
\oper{S}_z=\frac 1 2\hbar\(\ket{v}\bra{v}-\ket{u}\bra{u}\),
\end{equation}
where $\ket{u}$ is orthogonal on $\ket{v}$.
Such an operator always exists, and since $\ket{v}$ determines $\ket{u}$ up to a phase factor, it is unique.
Moreover, since the spin's particle was never observed before, it wasn't revealed yet by the previous observations.
Therefore, we can choose the spin operator $\oper{S}_z$ freely, so that the observation of the spin along the $z$-axis yields a definite result without requiring collapse.
But we can also choose $\oper{S}_z$ to be the $\frac 1 2\hbar\(\ket{u}\bra{u}-\ket{v}\bra{v}\)$, so that the definite outcome is the other possible one.
And we can do this for any observable, not only for spin.
We arrive at the following observation:
\begin{fquest}
\begin{observation}
\label{obs:fresh-observation}
When a system is observed for the first time, it is possible to assign the operator representing the observed property so that the observation happens unitarily, without having to appeal to collapse or branching.
\end{observation}
\end{fquest}

The situation is more complicated if incompatible properties of the same system $S$ are observed. For example, the system $S$ may be prepared in a state $\ket{\psi(0)}_S$ at the time $t=0$, and found to be in a state $\ket{\psi(T)}_S$ at the time $t=T$, $T > 0$.
Let $A$ and $B$ be two physical properties of the system $S$. We assume that the system $S$ is prepared so that its property $A$ has a definite value $a$, and when it is measured, its property $B$ has a definite value $b$.
We need to find what Hermitian operators $\oper{A}$ and $\oper{B}$, acting on the state space $\hilbert_S$ of $S$, represent the properties $A$ and $B$. Is it possible that $S$ is prepared to be in an eigenstate represented by an eigenvector $\ket{\psi(0)}_S$ of $\oper{A}$, with the eigenvalue $a$, and it is found to be in an eigenstate represented by an eigenvector $\ket{\psi(T)}_S$ of $\oper{B}$, with the eigenvalue $b$?

If this was possible, and if the property $A$ is conserved from $t=0$ to $t=T$, it seems that the system $S$ should have, at $t=T$, definite values for both $A$ and $B$.
But this is in general not possible. For example, it cannot have definite spins along both the $z$-axis and the $x$-axis.
It seems that we are still forced to choose between a jump of the state of $S$ from $\oper{U}_S(T,0)\ket{\psi(0)}_S$ to $\ket{\psi(T)}_S$ (wavefunction collapse), or a branching (as in MWI or PWT).

But, as we will see, this is not necessary. It only seems so, because we believe that we observe directly the system $S$, both when we prepare it so that property $A$ has a definite value, and when we measure its property $B$.
But what we observe in fact are the macroscopic properties of the pointer of the device making the preparation and of the pointer of the measuring device. By taking this into account, we can both maintain unitary evolution and avoid branching.

In the usual formulation of quantum mechanics, in which each physical property is represented by a pre-assigned operator, this can be achieved by allowing the environment change the state of the observed system between the preparation and the observation, so that the measurement has a definite outcome \citep{Schulman1984DefiniteMeasurementsAndDeterministicQuantumEvolution,Schulman2017ProgramSpecialState,Stoica2008SmoothQuantumMechanics,Stoica2012QMGlobalAndLocalAspectsOfCausalityInQuantumMechanics,Stoica2016OnTheWavefunctionCollapse,Stoica2017TheUniverseRemembersNoWavefunctionCollapse,Stoica2021PostDeterminedBlockUniverse}.
But such a solution requires very fine-tuned initial conditions for the state vector of the world, and this fine tuning has to take into account what measurements will be done throughout the history of the world. The initial state of the entire world should be arranged so that, at any measurement, the interactions between the observed system and the environment perturbs the observed system in the exact way needed to be found in a definite eigenstate.

The belief that the observation of a property $B$ for a system prepared so that another property $A$ has a definite value requires collapse may seem to be totally warranted, but it is not.
We don't have $100\%$ control of what's going on during the two main interactions required for the preparation and for the observation of the system, and also of the possible interactions with the environment between these two events.
We can only know the macroscopic properties of the preparation device, the measuring device, and the environment in general.
Since these systems consist of a huge number of particles whose microstates are practically unknown to us, there are an immense number of unknown variables, and they are totally out of our control.

In the standard paradigm of absolute assignment of operators to the physical properties, a disturbance, a ``kick'' as Schulman calls it \citep{schulman1991definiteQuantumMeasurements}, can make the system get into any allowed eigenstate by unitary evolution alone. A version of the idea was rediscovered by Stoica, who proposed in \citep{Stoica2008SmoothQuantumMechanics} that the state of the total system can be seen, from the point of view of someone applying standard quantum mechanics, as an entangled state between the observed system and the preparation device, and when the observation is finalized, we can select retroactively a separate state from this entangled state, so that the definite outcome is obtained without branching or collapse. But it could be that the system was all this time in the right state to get a definite outcome, so that the entanglement was only due to our description and the past is not ``rewritten''. The idea is illustrated in Figure \ref{fig:conspiracy}.

Suppose the observed system $S$ is prepared in the state $\textcolor{blue}{\ket{\psi_0}_S}\in\hilbert_S$. After a time interval $T$, we expect the observed system to be in the state $\textcolor{red}{\ket{\psi}_S}$, but our observation finds it in the state $\textcolor{blue}{\ket{+}_S}$. The other possible outcome, which was not obtained, is $\textcolor{blue}{\ket{-}_S}$, so that $\textcolor{red}{\ket{\psi}_S}=\textcolor{blue}{a\ket{+}_S}+\textcolor{blue}{b\ket{-}_S}$, where $a,b\in\C$, $\abs{a}^2+\abs{b}^2=1$.

\begin{ffig}
\begin{center}
\begin{tikzpicture}[scale=1.75]
	\draw[draw=gray!50!cyan!50!white, dashed, pattern=north west lines, pattern color=gray!50!cyan!50!white] (-0.75,-0.125) rectangle ++(1.5,3.5);
	\node at (0,1.4) {disturbance};
	\node at (-0.75, -0.65) {preparation};
	\node at (4.5, -0.65) {observation};

  \draw[->, >=stealth, line width=0.2mm] (-2, -0.25) -- (6, -0.25) node[above left] {time};
  \draw (-0.75,-0.3) -- (-0.75,-0.2) node [below]  {$t=0$};
  \draw (+0.75,-0.3) -- (+0.75,-0.2) node [below]  {$t=\varepsilon$};
  \draw (4.5,-0.3) -- (4.5,-0.2) node [below]  {$t=T$};

  \draw[->, >=stealth, line width=0.2mm] (-1.25, -0.125) -- (-1.25, 1.5) node[below left] {$\hilbert_S$};
  \draw[->, >=stealth, line width=0.2mm] (-1.25, 1.75) -- (-1.25, 3.5) node[below left] {$\hilbert_E$};

  \draw[domain=-2:-1, smooth, variable=\x, line width=0.25mm, blue] plot ({\x}, {0.75});
  \draw[domain=-1:1, smooth, variable=\x, line width=0.2mm, blue] plot ({\x}, {0.875+0.125*tanh(3.5*\x)});
  \draw[domain=-2:2, smooth, variable=\x, line width=0.2mm, blue, dashed] plot ({0.5*\x}, {0.5-0.25*tanh(1.5*\x)});
  \draw[domain=-2:5.5, smooth, variable=\x, line width=0.2mm, red, dotted] plot ({\x}, {0.75});
  \draw[domain=1:5.5, smooth, variable=\x, line width=0.2mm, blue] plot ({\x}, {1.0});
  \draw[domain=1:5.5, smooth, variable=\x, line width=0.2mm, blue, dashed] plot ({\x}, {0.25});
	
\draw[blue,fill=blue] (-0.75,0.125) circle (.25ex);
\draw[blue,fill=blue] (-0.75,0.75) circle (.25ex) node[above,blue] {$\ket{\psi_0}_S$};
\draw[blue,fill=blue] (4.5,0.25) circle (.25ex) node[below,blue] {$\ket{-}_S$};
\draw[blue,fill=blue] (4.5,1.0) circle (.25ex) node[above,blue] {$\ket{+}_S$};
\draw[red,fill=red] (4.5,0.75) circle (.25ex) node[below,red] {$\ket{\psi}_S$};

  \draw[domain=-2:-1, smooth, variable=\x, line width=0.2mm, violet, dashed] plot ({\x}, {2.5});
  \draw[domain=-2:-1, smooth, variable=\x, line width=0.2mm, violet] plot ({\x}, {2.125});
  \draw[domain=-1:1, smooth, variable=\x, line width=0.2mm, violet] plot ({\x}, {2.0-0.125*tanh(3.5*\x)});
  \draw[domain=-2:2, smooth, variable=\x, line width=0.2mm, violet, dashed] plot ({0.5*\x}, {2.75+0.25*tanh(1.5*\x)});
  \draw[domain=1:5.5, smooth, variable=\x, line width=0.2mm, violet, dashed] plot ({\x}, {3.0});
  \draw[domain=1:5.5, smooth, variable=\x, line width=0.2mm, violet] plot ({\x}, {1.875});

\draw[violet,fill=violet] (4.5,1.875) circle (.25ex) node[above,violet] {$\ket{+}_E$};
\draw[violet,fill=violet] (4.5,3.0) circle (.25ex) node[above,violet] {$\ket{-}_E$};
\draw[violet,fill=violet] (-0.75,2.125) circle (.25ex) node[below,violet] {$\ket{+}_{0E}$};
\draw[violet,fill=violet] (-0.75,2.5) circle (.25ex) node[above,violet] {$\ket{-}_{0E}$};
\end{tikzpicture}

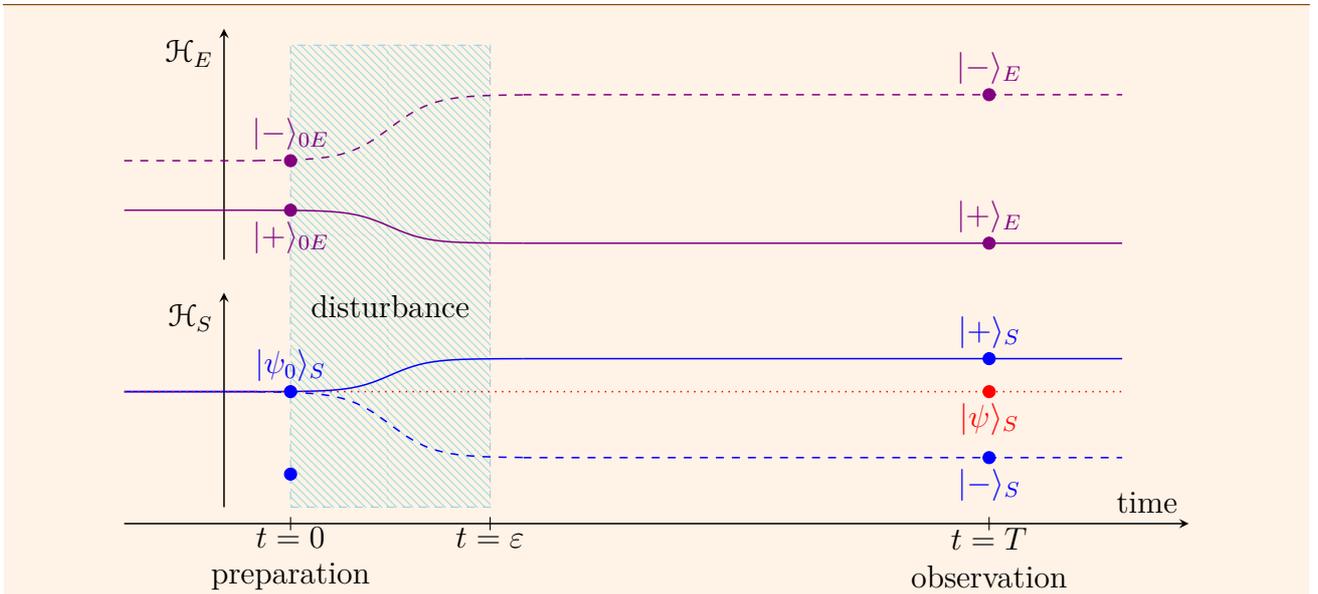
\captionof{figure}{An interaction between the observed system $S$ and the environment $E$ can disturb the observed system, so that, at the time $t=T$, instead of the evolved state $\textcolor{red}{\ket{\psi}_S}$, we get one of the observed outcomes $\textcolor{blue}{\ket{+}_S}$ or $\textcolor{blue}{\ket{-}_S}$. This would allow us to avoid collapse and branching.}
\label{fig:conspiracy}
\end{center}
\end{ffig}

A definite outcome could be obtained if an unobservable interaction happened, say during the time interval $(0,\varepsilon)$, disturbing the observed system to put it on the right track to evolve into $\ket{+}_S$ at $t=T$.
That part of the environment (in which we can include the preparation device, or even the measuring device if $\varepsilon\approx T$) that interacted with the observed system is also affected, so that, at $t=T$, its state becomes $\textcolor{violet}{\ket{+}_E}\in\hilbert_E$.
This disturbance of the observed system can be the kick proposed by Schulman.
The alternative outcome could be obtained by another disturbance, corresponding to the evolutions represented by dashed lines in Figure \ref{fig:conspiracy}.

We can see this as an entangled state $a\ket{+}_S\ket{+}_E+b\ket{-}_S\ket{-}_E$, in which the two outcomes are correlated with changes in the environment, but we can select out of this entanglement a single product state, for example $\textcolor{blue}{\ket{+}_S}\textcolor{violet}{\ket{+}_E}$, as in Figure \ref{fig:conspiracy} \citep{Stoica2008SmoothQuantumMechanics}.

As explained in \citep{Stoica2008SmoothQuantumMechanics}, this would also ensure the conservation laws, normally violated by the wavefunction collapse \citep{Burgos1993ConservationLawsQM,Stoica2017TheUniverseRemembersNoWavefunctionCollapse,Stoica2021PostDeterminedBlockUniverse}.
Moreover, in some cases there is a reciprocity: the restoration of the conservation laws also restores unitarity, as shown in \citep{Stoica2017TheUniverseRemembersNoWavefunctionCollapse}.
To restore the conservation laws in standard quantum mechanics, Collins and Popescu proposed a model based on entanglement \citep{CollinsPopescu2024ConservationLawsForEveryQuantumMeasurementOutcome} similar to this model proposed by Stoica in \citep{Stoica2008SmoothQuantumMechanics}.

This suggest the following possibility to update standard quantum mechanics to avoid collapse:
\begin{fquest}
\begin{observation}[Uncontrollable microscopic variables]
\label{obs:observation-uncontrollable}
The unknown and uncontrollable microscopic variables characterizing the preparation device, the measuring device, and the environment between them, may happen to have the right values so that these systems disturb the observed system, prepared in an eigenstate of the operator $\oper{A}$, with the exact amount needed to make it end out in an eigenstate of the operator $\oper{B}$.
\end{observation}
\end{fquest}

Of course, without a reason to constrain the initial conditions of all possible states specifically to ensure the existence of the appropriate disturbance, such a fluke would be extremely unlikely, having zero probability to happen by chance if the full state space is available for the initial conditions \citep{Stoica2012QMQuantumMeasurementAndInitialConditions}.
However, there is an alternative: if there are enough physical properties unassigned yet to operators, the freedom in assigning them can allow a single outcome by unitary evolution alone, and without such a conspiracy.
And there is a huge number of microscopic variables in the environment that are still unassigned to physical properties before the observation.

This can also account for more complicated experiments, like the EPR-Bell experiment \citep{EPR35,Bohm1951TheParadoxOfEinsteinRosenAndPodolsky,Bell64BellTheorem}.
This experiment consists of a preparation in the singlet state, so that the two entangled particles propagate to two different locations, and a joint measurement of two spin observables $\oper{S}_A$ and $\oper{S}_B$, taking place at the two different locations, on the two different particles.
Therefore, the EPR-Bell is a particular case of a preparation followed by a measurement.
The fact that these measurements happen at different locations is irrelevant, since what is measured is the joint observable $\oper{S}_A\otimes\oper{S}_B$.
Suppose a pair of spin-$1/2$ particles is prepared in a singlet state $\frac{1}{\sqrt{2}}\(\ket{\uparrow}_A\otimes\ket{\downarrow}_B-\ket{\downarrow}_A\otimes\ket{\uparrow}_B\)$, and Alice and Bob measure the spin along the $z$-axis.
The interaction with the device that prepares the observed system to be a singlet state as in Figure \ref{fig:conspiracy}, may in fact result in a product state, $\ket{\uparrow}_A\otimes\ket{\downarrow}_B$ or $\ket{\downarrow}_A\otimes\ket{\uparrow}_B$, so that the outcomes of the measurements done by Alice and Bob are anti-correlated as expected.
If Bob measures the spin along another axis at $45^\circ$ with respect to the $z$-axis, the interaction with the preparation device leaves the observed system in one of the separate states $\ket{\uparrow}_A\otimes\ket{\nearrow}_B$, $\ket{\downarrow}_A\otimes\ket{\nearrow}_B$, $\ket{\uparrow}_A\otimes\ket{\swarrow}_B$, or $\ket{\downarrow}_A\otimes\ket{\swarrow}_B$.
Again, such a conspiratorial interaction leading unitarily to definite outcomes for Alice and Bob's observations is possible in principle, on the behalf of the interaction with the preparation device. And again, the conspiracy can be removed by adopting the explanation of the delayed assignment of what operators represent the spins along the two axes.

From this non-quantitative explanation it follows that
\begin{fquest}
\begin{observation}[Observation as physication]
\label{obs:observation-physication}
The delayed assignment of operators to physical properties allows the observed system to evolve unitarily into the observed state, without wavefunction collapse or branching, and without a special choice of the initial conditions.
\end{observation}
\end{fquest}

Such a non-conspiratorial solution can be obtained from a conspiratorial unitary solution in the following way.
Suppose we think that the system is in a state that leads, after the measurement, to a superposition of indefinite outcomes.
Then, I will argue that there is a unitary transformation that ``corrects'' the presumed state of the system into a state that achieves a unique outcome without collapse.
Given the huge number of variables defining the total state space, compared to the number of macroscopic variables, there must exist numerous unitary transformations of the total state space that can change microscopic variables that are not macroscopic, but not the macroscopic ones.
In the standard formulation of quantum mechanics, where the assignment is assumed to preexist, the possibility of a solution is based on the same observation that the values of most microscopic degrees of freedom are unknown.
Since the conspiratorial solution achieves a definite outcome by unitary evolution alone, the same is true of the one obtained by delayed assignment, which in addition is non-conspiratorial.
The conspiracy is avoided by delaying the assignment of some microscopic physical properties. If all assignments were made at the beginning of the universe, this solution wouldn't work.
This is the main difference between the interpretation of quantum mechanics proposed here and the other interpretations, including standard quantum mechanics.

\section{Postulates}
\label{s:physication-postulates}

In this Section I present the Postulates of the new interpretation sketched in the previous Section.
The first three Postulates and the fifth Postulate of the proposed interpretation are common features of standard quantum mechanics, but they should be understood in a situation where not all physical properties have a priori assigned operators, and the assignment happens in time. This delayed assignment will be relevant in Postulate \ref{pp:assignment}, which replaces the Projection Postulate.

I take as a first postulate that there is a state vector undergoing a unitary evolution.

\begin{frem}
\begin{postulate}[Unitary evolution]
\label{pp:unitary}
The state of the universe is represented by an abstract unit vector $\ket{\psi(t)}$ in a complex \emph{state space} $\hilbert$. The \emph{evolution law} is given by the Schr\"odinger equation
\begin{equation}
	\label{eq:unitary_evolution}
	\ket{\psi(t)}=\oper{U}(t,t_0)\ket{\psi(t_0)},
\end{equation}
where $\oper{U}(t,t_0):=e^{-i/\hbar(t-t_0)\oper{H}}$ and $\oper{H}$ is the Hamiltonian operator. 
\end{postulate}
\end{frem}

It is important to emphasize that
\begin{enumerate}
	\item 
$\ket{\psi(t)}$ is just an abstract unit vector, and all unit vectors are identical, in the sense that there is a unitary transformation that maps any unit vector into any other unit vector,
	\item 
I don't assume an assignment of operators to physical properties at this explanatory stage, but this will be specified by the other Postulates.
\end{enumerate}

In particular, no interpretation $\psi(\xThree_1,\xThree_2,\ldots)=\braket{\xThree_1,\xThree_2,\ldots}{\psi(t)}$ of the unit vector $\ket{\psi(t)}$ as a wavefunction, where $(\xThree_1,\xThree_2,\ldots)$ are points in the configuration space of the classical theory from which our quantum theory may be obtained by quantization, is given at this point.

The operator $\oper{H}$ is not a matrix $H_{jk}:=\bra{\xThree_1^j,\xThree_2^j,\ldots}\oper{H}\ket{\xThree_1^k,\xThree_2^k,\ldots}$, where $(\xThree_1^j,\xThree_2^j,\ldots)$ and $(\xThree_1^k,\xThree_2^k,\ldots)$ are points in some configuration space. The operator $\oper{H}$ is just an abstract operator. The same is true of the unitary evolution operators $\oper{U}(t,t_0):=e^{-i/\hbar(t-t_0)\oper{H}}$.

However, the Hamiltonian operator $\oper{H}$ can be expressed in terms of some other operators supposed to represent physical properties like positions, similar to equation \eqref{eq:schrod_hamiltonian_NRQM}, but there is no a priori assignment of these other operators, and therefore no a priori configuration or phase space. The assignment will be done gradually, as observations happen.
But what happens in reality is that even this dependency of the Hamiltonian operator on other operators, like the one from equation \eqref{eq:schrod_hamiltonian_NRQM}, follows from observations.
Therefore, I will postulate the existence of such a functional dependence of the Hamiltonian operator on other operators, which have specific spectral properties, as we often do, but it is not necessary to postulate that the functional form of this dependence is already known to us.

\begin{frem}
\begin{postulate}[Structure of dynamical law]
\label{pp:structure-dyn}
There is a set of Hermitian operators $\{\oper{a}_1,\oper{a}_2,\ldots\}$ on $\hilbert$, so that the Hamiltonian operator $\oper{H}$ depends functionally on these operators,
\begin{equation}
\label{eq:Hamiltonian-structure}
\oper{H}=\ms{H}\(\oper{a}_1,\oper{a}_2,\ldots\).
\end{equation}

The operators $\{\oper{a}_1,\oper{a}_2,\ldots\}$ are characterized by their spectra, including multiplicity, and they  represent some physical properties $\{a_1,a_2,\ldots\}$, but the assignment
\begin{equation}
\label{eq:assignment-variables}
a_j \mapsto \oper{a}_j,
\end{equation}
is not known a priori, it will gradually result from observations.
\end{postulate}
\end{frem}

The operators $\{\oper{a}_1,\oper{a}_2,\ldots\}$ may be given only up to a unitary transformation. For example, they can be position-like and momentum-like operators $\{\hat{\xThree}_1,\hat{\pThree}_1,\hat{\xThree}_2,\hat{\pThree}_2,\ldots\}$, satisfying the CCR, and the dependency equation \eqref{eq:Hamiltonian-structure} can be equation \eqref{eq:schrod_hamiltonian_NRQM}.
But, for generality, I will not assume that these operators are necessarily position and momentum operators, they can be field operators, including gravitational field operators from an yet to be found quantum gravity.

The operators $\{\oper{a}_1,\oper{a}_2,\ldots\}$ may apply to subsystems, but they will be assumed to apply to the entire world. That is, if the operator of interest $\oper{a}_j^\alpha$ is defined on a state space $\hilbert_\alpha$ of a subsystem, where
\begin{equation}
\label{eq:TPS-alpha}
\hilbert\cong\hilbert_\alpha\otimes\hilbert_{\tn{rest of the world}},
\end{equation}
we will take the operator $\oper{a}_j:=\oper{a}_j^\alpha\otimes\oper{I}_{\tn{rest of the world}}$, defined on the total state space $\hilbert$, instead of the operator $\oper{a}_j^\alpha$ defined on $\hilbert_\alpha$.

Moreover, the \emph{tensor product structure} (TPS), the decomposition of the state space as a tensor product of state spaces representing the subsystems, for example the one from equation \eqref{eq:TPS-alpha}, is also not given a priori and should follow from experiments. The TPS is given by a set of algebras of operators $\mc{A}_j$, $j\in\{1,2,\ldots\}$, so that the elements of an algebra $\mc{A}_j$ commute with the elements of another algebra $\mc{A}_k$, $j\neq k$ \citep{ZanardiLidarLloyd2004QuantumTensorProductStructuresAreObservableInduced}.
These algebras of operators are not assigned a priori to physical properties of subsystems, so the TPS itself is not given a priori\footnote{Some authors claim that, under conditions like locality, the abstract Hamiltonian (or its spectrum) determines a unique TPS \citep{CotlerEtAl2019LocalityFromSpectrum}, but this was disproved in \citep{Stoica2022SpaceThePreferredBasisCannotUniquelyEmergeFromTheQuantumStructure,Stoica2024DoesTheHamiltonianDetermineTheTPSAndThe3dSpace}. Already for finite-dimensional state spaces, regardless of the conditions that we impose to the TPS, if there is at least a TPS satisfying those or any other conditions, the number of additional parameters that need to be fixed to specify a unique TPS satisfying the same conditions grows exponentially with the number of factors in the TPS \citep{Stoica2024DoesTheHamiltonianDetermineTheTPSAndThe3dSpace}.}.

The unitary evolution law is very general, and a theory of the physical world should be much more specific, but we don't need to discover here the actual Hamiltonian of the world.
Postulate \ref{pp:structure-dyn} only states that there must be some operators and the Hamiltonian must depend on them in an yet unknown way, but it leaves open the question of which way this is. This should follow as well from experimental observations and future theoretical developments, just like in standard quantum theory and any other interpretation.
But it is important to mention the existence of these operators and the functional dependency of the Hamiltonian of these operators.


But an observation requires a measuring device and a system to be observed with the help of the measuring device.
Both of these are subsystems, and so they have to be represented on some state spaces $\hilbert_M$ (for the measuring device) and $\hilbert_S$ (for the observed system). These state spaces have to be factors in a tensor product structure of the total state space $\hilbert$,
\begin{equation}
\label{eq:TPS-M-S}
\hilbert\cong\hilbert_S\otimes\hilbert_M\otimes\hilbert_{\tn{rest of the world}}.
\end{equation}

In the following, I will rely again on the undeniable fact that the TPS can be obtained from the operators, more or less along the lines from \citep{ZanardiLidarLloyd2004QuantumTensorProductStructuresAreObservableInduced}.
Therefore, I will take these operators as acting on the entire state space $\hilbert$, not restricted to factors like $\hilbert_S$ or $\hilbert_M$. For example, an operator $\oper{A}_S$ on $\hilbert_S$ can be identified with an operator $\oper{A}$ on $\hilbert$ by tensoring it with the identity operators $\oper{I}$ of the other factors from equation \eqref{eq:TPS-M-S},
\begin{equation}
\label{eq:oper-extended}
\oper{A}:=\oper{A}_S\otimes\oper{I}_M\otimes\oper{I}_{\tn{rest of the world}}.
\end{equation}

Therefore, the decomposition \eqref{eq:oper-extended} depends as well on the assignment of operators to the physical properties, and it is not completely determined by the observations, which reveal only the assignments of the macroscopic physical properties.

The word ``macroscopic'' is not strictly about size:
\begin{fquote}
\begin{definition}
\label{def:macro}
A \emph{macroscopic physical property} is a property that can be observed without doing quantum experiments, and it always has a definite value. 
\end{definition}
\end{fquote}

But let's keep in mind that, even though our universe is fully quantum, not all observations consist of quantum experiments done in a laboratory. In fact, most of them are direct observations done at the macroscopic level (that is, they don't require quantum experiments to be observed). Most of us never did an explicitly quantum experiment, and humanity already gathered much knowledge before the advent of quantum mechanics.
Moreover, to even build a measuring device, we need to rely on such direct knowledge of macroscopic properties.
In fact, even when we do a quantum measurement, we make only direct macroscopic observations:
\begin{enumerate}
	\item 
We make sure, based on our macroscopic experience, that the measuring device and the device that prepares the system to be observed are built and work properly, and that they are set up properly.
	\item 
We observe directly the pointer state of the measuring device, and our only knowledge of the observed system's state is inferred from the macroscopic properties of the whole set-up and the pointer state.
\end{enumerate}

Whatever knowledge we gather about the quantum world, we do it by direct reading of macroscopic physical properties, combined with inference. Therefore,

\begin{fquest}
\begin{observation}
\label{obs:quantum-measurement}
Quantum measurements are a particular case of macroscopic observations.
\end{observation}
\end{fquest}

That is, we never observe the microscopic properties that we think we measured, we only infer them from the macroscopic ones. The physical properties available for observation include the pointer states of the measuring devices, the records of outcomes of past experiments, and everything else that is available to us by direct observation of the macroscopic world.

As in the case of Postulate \ref{pp:structure-dyn}, I will assume that the assignment of operators to the macroscopic physical properties doesn't exist a priori.
Each of the properties $M_j$ is represented by a Hermitian operator $\oper{M}_j$, and can be expressed by a numerical value which is an eigenvalue of $\oper{M}_j$, but we don't know a priori what are these operators.
All macroscopic physical properties are compatible, so they should be represented by some commuting operators $\oper{M}_1,\oper{M}_2,\ldots$. 

\begin{frem}
\begin{postulate}[Macroscopic physical properties]
\label{pp:macro-properties}
There is a set
\begin{equation}
\label{eq:macro-properties}
\mc{M}:=\{M_1,M_2,\ldots\}
\end{equation}
of compatible macroscopic physical properties, in principle available for direct observation.
There is an assignment of operators to the macroscopic physical properties,
\begin{equation}
\label{eq:assignment-macro}
M_j \mapsto \oper{M}_j,
\end{equation}
so that, for all $j\neq k$,
\begin{equation}
\label{eq:assignment-macro-commute}
\oper{M}_j\oper{M}_k=\oper{M}_k\oper{M}_j,
\end{equation}
but this assignment is not a priori, it happens gradually in time.
\end{postulate}
\end{frem}

Let us see how physication works to solve the measurement problem.

There is a unique decomposition of the state space $\hilbert$ as a direct sum of maximal subspaces consisting of common eigenvectors of the operators $\oper{M}_1,\oper{M}_2,\ldots$,
\begin{equation}
\label{eq:macrostates-decomposition}
\hilbert=\bigoplus_{a\in\ms{A}}\hilbert_a,
\end{equation}
where $\ms{A}$ is a set of labels of the subspaces.
Each subspace $\hilbert_a$ consists of common eigenvectors of the operators $\oper{M}_1,\oper{M}_2,\ldots$, so that for each operator $\oper{M}_j$, all vectors from $\hilbert_a$ are eigenvectors for the same eigenvalue $\lambda_j(a)$.

We call the subspaces $\hilbert_a$ from the decomposition \eqref{eq:macrostates-decomposition} \emph{macrostates}. To each subspace $\hilbert_a$ we associate a \emph{macroprojector}, \ie a projection operator $\oper{P}_a$ so that
\begin{equation}
\label{eq:macrostates}
\hilbert_a=\oper{P}_a\hilbert.
\end{equation}

Equation \eqref{eq:macrostates} implies that each macroscopic operator is of the form
\begin{equation}
\label{eq:macroscopic-operator-form}
\oper{M}_j=\sum_{a\in\ms{A}}\lambda_j(a)\oper{P}_a.
\end{equation}
Here the sum symbol is formal, the decomposition can consist of integrals as well, if the spectrum of the operators representing macroscopic physical properties has continuous parts.

Let us consider now the initial state $\ket{\psi_0}$ of the universe at the Big Bang. According to the Past Hypothesis \ref{obs:past-hypothesis}, this state had minimal entropy and its macroscopic properties were well-defined.
That is, it belonged to an initial macrostate of low entropy, let it be $\hilbert_0$,
\begin{equation}
\label{eq:state-init}
\ket{\psi_0}\in\hilbert_0.
\end{equation}

As the time goes, the entropy of the state represented by $\ket{\psi(t)}$ increases. That is, for times $t_2> t_1>t_0$, the vectors from the subspace $\oper{U}(t_2,t_0)\hilbert_0$ represent states of higher entropy than those from the subspace $\oper{U}(t_1,t_0)\hilbert_0$.
Each subspace $\oper{U}(t,t_0)\hilbert_0$ can be decomposed in terms of macrostates,
\begin{equation}
\label{eq:evol-macro-decomposition}
\oper{U}(t,t_0)\hilbert_0=\bigoplus_{a\in\ms{A}(t)}\hilbert_a,
\end{equation}
where $\ms{A}(t)\subset\ms{A}$ is a subset of the set of labels $\ms{A}$ used in the decomposition \eqref{eq:macrostates-decomposition}.
For example, equation \eqref{eq:state-init} states that $\ms{A}(t_0)=\{0\}$.

Since for different times $t_2>t_1>t_0$ the entropy is different, it follows that there are macroscopic observables that are different in the decompositions \eqref{eq:evol-macro-decomposition} for different times, so that the macrostates are different too, and
\begin{equation}
\label{eq:evol-macro-distinct}
\oper{U}(t_1,t_0)\hilbert_0\perp\oper{U}(t_2,t_0)\hilbert_0.
\end{equation}

Therefore, 
\begin{equation}
\label{eq:evol-macro-distinct-labels}
\ms{A}(t_1)\cap\ms{A}(t_2)=\varnothing.
\end{equation}

As $\ket{\psi(t)}$ evolves in time, it can became an eigenstate of some of the operators representing macroscopic properties, the eigenvalues of the macroscopic operators may change, or $\ket{\psi(t)}$ can become a superposition of such eigenstates.
However, since most physical properties don't have assigned operators, this leaves us a freedom. 
Equation \eqref{eq:state-init} fixes the assignment for the subspace $\hilbert_0$. As soon as $\ket{\psi(t)}$ leaves the subspace $\hilbert_0$, say at $t_1$, it belongs to a subspace $\oper{U}(t_1,t_0)\hilbert_0$, which is a direct sum of other subspaces $\bigoplus_{a\in\ms{A}(t_1)}\hilbert_a$. But since these subspaces are not assigned yet to physical properties, they can be assigned at $t_1$ so that $\ket{\psi(t_1)}$ belongs to only one of them!
This means that the simple fact that there is change all the time implies that
\begin{fquest}
\begin{observation}
\label{obs:macro-assignment-freedom}
There is a freedom to choose the assignment of operators to the macroscopic physical properties so that $\ms{A}(t)$ always has only one element.
\end{observation}
\end{fquest}

As explained in Section \sref{s:physication}, if the initial state $\ket{\psi_0}$ is not predetermined, it is possible to choose it so that quantum measurements have definite outcomes. In our case, $\ket{\psi_0}$ is given a priori, but most of the microscopic physical properties are not pre-assigned. And it turns out that the freedom to choose the microscopic physical properties that are not assigned a priory is the same as the freedom to choose $\ket{\psi_0}$ so that quantum measurements have definite outcomes.

This leads us to the Postulate where physication kicks in:
\begin{frem}
\begin{postulate}[Assignment]
\label{pp:assignment}
Whenever $\ket{\psi(t)}$ leaves the already assigned macrostates, the new ones are assigned so that $\ket{\psi(t)}$ is in a definite macrostate.
\end{postulate}
\end{frem}

That is, there is an element $a(t)\in\ms{A}$ so that
\begin{equation}
\label{eq:assignment-psi}
\ket{\psi(t)}\in\hilbert_{a(t)}.
\end{equation}

This accounts for definite outcomes, resolving the measurement problem, at least for non-relativistic quantum mechanics or, in general for quantum theory with an absolute decomposition of spacetime into space and time.
But it is also consistent with the relativity of simultaneity, since the order of assignment should not be taken as absolute. For example, for the EPR-Bell experiment, Alice and Bob do local measurements at spacelike separated locations, so the assignments they make are local and spacelike separated. So the order of assignment will depend on the observer, an observer may see Alice doing the measurement before Bob does it, while another observer may see Bob doing the measurement first. All that matters is that the assignments they do are consistent with one another.

Based on the past history of the system we may anticipate that $\psi(t)$ will become a superposition of macrostates
\begin{equation}
\label{eq:pre-assignment-psi}
\ket{\psi(t)}=\sum_{a\in\ms{A}(t)}\oper{P}_a\ket{\psi(t)},
\end{equation}
but Postulate \ref{pp:assignment} states that the assignment happens so that $\oper{P}_a\ket{\psi(t)}=0$ for all $a\in\ms{A}(t)$ except for one of them, denoted by $a(t)$.

\begin{fquote}
\begin{definition}
\label{def:physication}
We call the process of assignment \emph{physication}.
\end{definition}
\end{fquote}

Recall the dilemma forced to us by Bell's theorem \citep{Bell2004SpeakableUnspeakable}, which some interpret as a trilemma: that we have to either give up locality, or to embrace conspiracy (Bell calls it ``superdeterminism''), or to embrace many worlds.
Does physication avoid all these three undesirable features?
It does so only by recognizing that there is no pre-assignment of operators to the physical properties. This assignment happens gradually, with each new observation, and it has to be consistent, so that it doesn't contradict previous assignment.
This raises the question: how then can two observers at a distance contribute to physication, in agreement with Postulate \ref{pp:assignment}? Postulate \ref{pp:assignment} mentions only macrostates, but isn't this hiding under the rug some non-locality?
For the EPR-Bell experiment, the assignments done by Alice and Bob have to be consistent with one another. This means that the assignment, the physication itself, has to be non-local in some sense.
But it happens without non-local interaction, and once the assignment is done, it describes a physical world without non-local interactions.
Since the assignment is related to sentience \citep{Stoica2023AskingPhysicsAboutPhysicalismZombiesAndConsciousness}, does this indicate a sentiential relation between Alice and Bob?
Maybe this is related to another result, that mental states, in order to be integrated into the unified minds that each of us experiences, already require a sort of non-locality even for a single observer \citep{Stoica2020AreMentalStatesNonlocal}.
But this possible connection is not the subject of the present proposal.

Returning to our postulates, since the choice of assignment is not determined, we also need to specify the probability that $a(t)$ is one or another element of $\ms{A}(t)$.

\begin{frem}
\begin{postulate}[Probability]
\label{pp:probability}
The probability $P_a(t)$ that the macrostate at the time $t$ is $\hilbert_{a}$, where $a\in\ms{A}(t)$, is given by the \emph{Born rule}:
\begin{equation}
\label{eq:born-rule}
P_a(t)=\bra{\psi(t)}\oper{P}_a\ket{\psi(t)}.
\end{equation}
\end{postulate}
\end{frem}

It may be possible to derive the Born rule from the other Postulates, or maybe to derive different probabilities, violating the Born rule, and therefore invalidating this proposal.
But for the moment I state it as a separate Postulate, leaving the questions of independence and consistency of this Postulate with the previous ones for future research.

\section{Can we test it?}
\label{s:experiment}

The proposed interpretation relies on the existence in the environment of sufficiently many variables whose physical meaning can be assigned later, so that its interaction with the observed system leads to a definite outcome of the measurement. The more potentially influential particles are in the environment, in particular in the preparation device, the more possibilities to assign their parameters by an appropriate unitary transformation to achieve the result.
Therefore,
\begin{fquest}
\begin{observation}
\label{obs:ratio-free-vs-observed}
The possibility of unitary assignment grows, roughly, with the number of degrees of freedom with which the observed system interacted since the previous measurement.
\end{observation}
\end{fquest}

That is, if the preparation device is too small, the assignment can't be realized properly. But, on the other hand, if it's too small, it could also be the case that the preparation can't be done properly. A system can't act as a measuring device if it is too ``small'', in the sense of having too few degrees of freedom.
Smaller systems able to clone the state of another system with which they interact, in the sense that it acts like a measuring device, are possible. But such systems can count as actual measuring devices only if the degrees of freedom into which they ``copied'' the state are stable and can be observed macroscopically. Such macroscopic variables can only be aggregate variables of many particles, so that the action of the system is much larger than Planck's quantum of action $h$. This makes it so that the possibility of delayed assignment correlates with the possibility that the measuring device, in particular the preparation device, is large enough. Therefore, the interpretation proposed here predicts what we already know about the measuring devices, that they have to be large enough.

But maybe we can do better than merely predicting independently that measuring devices should consist of many degrees of freedom. 
Let us try some strategy to obtain empirical predictions, based on the possibility that, even if both measuring devices and delayed assignments require many degrees of freedom, the ranges of these degrees of freedom don't coincide.

The strategy can be to build a measuring device, to be used for preparations, whose functionality can be described based on the projection postulate, but not by delayed assignment.
If the experiments confirm the description based on the projection postulate, this would reject experimentally the interpretation proposed here.
If we can build such a device, and if its degrees of freedom that are not macroscopically observable are not sufficient to allow the assignments as described in Sections \sref{s:physication} and \sref{s:physication-postulates}, then this would reject the interpretation proposed here.

Another possibility to falsify this proposal is to derive a contradiction between Postulates \ref{pp:unitary}--\ref{pp:assignment} and the Born rule, given provisionally as Postulate \ref{eq:born-rule}.
If this proposal is correct, maybe Postulate \ref{eq:born-rule} can be derived from the others, assuming a uniform distribution of the initial states in $\hilbert_0$, as in equation \eqref{eq:state-init}.
But if the math will result in different probabilities than those given by the Born rule, this would be a refutation of this proposal, based on already known empirical data.

\section{Comparison with other proposals}
\label{s:comparison}

We have seen that the freedom of choosing the assignments is used in these Postulates to allow for a single-world unitary evolution.
When our current macrostate encodes information that the observed system was prepared in a particular way, if we apply the {\schrod} equation, we may infer that a measurement will find the system in a superposition.
But the macrostate of the world after the measurement indicates that the measurement has a definite outcome.
This is explained in Section \sref{s:physication-postulates} by the fact that the macrostate before and the one after the measurement are orthogonal, and this allows to assign the macroscopic properties in a way that accounts for both of these by the unitary evolution of the world. In Section \sref{s:physication}, this is interpreted microscopically as a spontaneous interaction with the preparation device, which seems conspiratorial if we think that all assignments are given a priori.

This proposal of an interpretation of quantum mechanics eliminates the necessity of collapse or branching of the wavefunction. There is a single world, and it is not conspiratorial, because the physical properties are assigned to operators as new observations are made. Had this assignment been done a priori, it would be conspiratorial, but by delaying it and leaving it to physication via the observations, the necessity of conspiracy is removed.

The central role of macroscopic physical properties, which are very similar to classical properties, may remind us of Bohr's views. Bohr considered that the measuring device is a classical system, and the outcomes of the measurements are seen as classical properties (\citealp{Stoica2022SpaceThePreferredBasisCannotUniquelyEmergeFromTheQuantumStructure}, \S7.1). The microscopic properties exhibit quantum effects, but these are not seen directly. 
The observed system and the measuring device should be taken into account as an undivided whole.
Indeed, if we take only the macroscopic properties as observable, the ``observed system'' is not actually observed, and so its properties are not macroscopic. But by taking it as forming an indivisible whole with the measuring device, the properties of the measuring device can tell us about the observed system, if we infer its properties from the macroscopic properties. This necessity to take the measuring device as the context of the observed system allows different experimental setups to measure complementary properties of the observed system.

In the Copenhagen Interpretation, Heisenberg saw the properties of the observed system as gaining reality only due to the measurement. This is also somewhat similar to physication, which allows us to assign operators to the physical properties of the observed system as a result of the observations we make.
Similar parallels can be made with neo-Copenhagen interpretations like QBism \citep{FuchsMerminSchack2014AnIntroductionToQBism}.

A possibility to interpret the views of Bohr, Heisenberg, and others, is not as creating reality, but merely as physication, \ie as assigning physical meaning to an existing but abstract reality.

According to Jaynes \citep{Jaynes1990ProbabilityInQuantumTheory},
\begin{fquote}
\begin{quote}
But our present QM formalism is not purely epistemological; it is a peculiar mixture describing in part realities of Nature, in part incomplete human information about Nature -- all scrambled up by Heisenberg and Bohr into an omelette that nobody has seen how to unscramble.
\end{quote}
\end{fquote}

I propose that the idea of physication can unscramble this omelette.
The unscrambling consist of distinguishing between structure, relational properties, which can be fully captured by mathematics, and their physical meaning, which can be acquired gradually, through observations.
Therefore, the unscrambling happens by replacing the epistemic aspect of quantum measurements with a mixed process, involving both an epistemic aspect and physication.

The structure, given by the state vector $\ket{\psi(t)}$ and the properties of macroscopic and other operators, is already given, like a mold, a matrix, while the physical properties, which cast the state vector into a wavefunction, are assigned as the observations take place, and this assignment applies to the past too.
For example, for non-relativistic quantum mechanics, position-like operators $\hat{\xThree}_1,\hat{\xThree}_2,\ldots$ are assigned to the physical positions, and they, together with the state vector, coalesce into the wavefunction $\braket{\xThree_1,\xThree_2,\ldots}{\psi(t)}$.
Since the state vector evolves towards the future, and the assignment of position-like operators ``propagates'' back in time to interpret the past of the system, this reminds us of another view, in which a state vector $\ket{\psi_p(t)}$ evolving from the past to the future is used together with the outcome of the measurement, a state vector $\bra{\psi_r(t)}$ which then from the future to the past. This is the \emph{two-state vector formalism} \citep{Watanabe1955SymmetryOfPhysicalLawsPartIIIPredictionAndRetrodiction,AharonovBergmannLebowitz1964TimeSymmetryInTheQuantumProcessOfMeasurement,AharonovVaidman2007TheTwoStateVectorFormalismAnUpdatedReview}.
In the two-state vector formalism there are two state vectors, which together allow us to describe what happens between the preparation and the measurement in a novel and seminal way, but the collapse still seems to be required for a third measurement. In the physication interpretation there is only one state vector, which is interpreted physically, for example as a wavefunction, starting from the outcome and going backwards in time, or, in the case of more complex experiments like the EPR-Bell experiment, zigzagging through spacetime. But only the assignment of physical meaning is interpreted as going backwards, the state vector simply evolves from the past to the future as usual.

As seen in our discussion of quantum measurements in terms of observations of the macroscopic pointers, when operators are assigned to the macroscopic physical properties, one can infer assignments to the microscopic physical properties as well. The assignments of macroscopic properties for what happened in the past remains unchanged, but the inferred assignments of microscopic properties are not absolute, they can be reevaluated when new observations are made. This is how physication avoids the necessity to invoke the wavefunction collapse.

If we forget this delayed assignment and assume it to be given a priori, we have the impression of conspiracy, superdeterminism, or retrocausality.
We understand that it is not conspiratorial only if we recall that the assignments are delayed and applied retroactively, but the past of the state vector $\ket{\psi(t)}$ itself is not changed retroactively.
Therefore, in the interpretation proposed here, all of these are illusions, there is no conspiracy.
And there is no collapse as in the collapse theories, there is only one history and not many as in MWI, and there are no additional hidden variables as in PWT.

\addcontentsline{toc}{section}{\refname}


\end{document}